\documentstyle[multicol,prl,aps,epsfig]{revtex}
\begin{document}
\draft
\title{Two-Order-Parameter Description of Liquids: \\
Critical Phenomena and Phase Separation of Supercooled Liquids}
\author{Hajime Tanaka}
\address{
Cavendish Laboratory, Madingley Road, Cambridge, CB3 0HE, UK. \\
Institute of Industrial Science, University of Tokyo, Minato-ku, Tokyo 106, 
Japan.\cite{A} 
}
\date{Received 17 June 1997}
\maketitle
\begin{abstract}
Because of the isotropic and disordered nature of liquids, 
the anisotropy hidden in intermolecular interactions 
are often neglected. 
Accordingly, the order parameter describing  
a simple liquid has so far been believed to be only density. 
In contrast to this common sense, we propose that two order parameters, 
namely, density and bond order parameters,  
are required to describe the phase behavior of 
liquids since they intrinsically tend to form local bonds.  
This model gives us clear physical explanations for 
two poorly-understood phenomena in supercooled liquids: 
(i) large-scale density fluctuations and (ii) phase separation of a 
one-component liquid into two liquid phases. 
\end{abstract}


\begin{multicols}{2}
Generally, a liquid phase is bounded by a gas phase and a solid phase,  
although some molecules having strong ``topological'' anisotropy exhibit 
additional phase known as a liquid crystalline phase above its solid 
phase \cite{Lubensky}. By lowering a temperature of a gas phase, 
the potential energy becomes more important than the kinetic 
energy and intermolecular interactions come into play. 
At a gas-liquid phase transition point, 
the attractive interaction leads to the formation of a ``liquid'' phase.
The principal physical quantity distinguishing  the liquid and the gas 
phases is their density.  
Both phases have complete translational and rotational symmetry, and 
they are isotropic and homogeneous. 
Because of these features, 
a liquid state is usually assumed to be described by 
only one order parameter, density. 

The physical properties of liquid are rather well understood 
at the temperature far from its stability limit, while  
very poorly understood especially near its liquid-solid 
phase transition \cite{Angell}. Here we mention three of such examples: 
(i) glass transition phenomenon itself 
\cite{Angell,Richert,Noncry}, (ii) 
large-scale density fluctuations in supercooled liquids 
\cite{Debye,Fischer2,Wang}, and (iii) 
an unusual phenomenon of phase separation 
of a simple one-component liquid into two liquid phases 
in a supercooled state \cite{Aasland,StanleyS,StanleyT}. 
The key to understand these phenomena, we believe, is 
to realize that the above common sense that 
a liquid state can be described by only density 
order parameter is basically ``wrong''. 
There are a number of studies that try to improve 
a theory of liquid by including many-body effects, or 
higher-order density correlations.  
The most important concept is the local bond orientaional 
order (LORO) \cite{Frank,LORO} based on the fact that even 
spherical molecules 
locally favor a tetraherdal configuration in a liquid phase: 
For example, Frank \cite{Frank} pointed out 
that icosahedral clusters of 13 particles have a 
significantly lower energy than more obvious ``crystallographic'' 
arrangements of the corresponding 
fcc or hcp structures. 
Based on this idea, he explained why a simple liquid metal can be 
so deeply supercooled. 
This concept of LORO leads to the recent development 
along this direction \cite{Frank,LORO,Cohen2,Glarum,Blender,Yeo}.  
However, we also need consider the anisotropic 
specific interactions that commonly exist 
in molecules having no spherical symmetry \cite{HTJCP}. 
The most extreme case is the network formation due to 
hydrogen bondings, as in SiO$_2$.    
For a general physical description of real liquids, thus,  
we need to introduce a new, nontrivial order parameter 
that can describe both topological and energetical effects  
inducing the locally favored configuration of liquid molecules 
or atoms and, then,  
to clarify the effects of its frustration with 
crystallization on the phase behavior of liquids.  

The thermodynamic state of condensed matter can be described 
solely by the balance between the energetic interactions 
and the entropy of the system. 
Thus, we here focus our attention on an intermolecular attractive 
potential. 
The attractive interaction potential between a molecule and its neighbors 
is generally given by the form $V(r,\Omega) 
=\bar{V}(r)+\Delta V(r,\Omega)$, 
where $r$ is the distance from the molecules 
and $\Omega$ expresses the orientation. 
$\bar{V}$ represents the isotropic part of   
the interaction, while $\Delta V$ its anisotropic part. 
This anisotropy comes from both (i) locally favored geometrical packing  
under attractive interactions \cite{Frank,LORO} and (ii) specific 
anisotropic interactions, as explained above.  
We believe that these two types of anisotropic interactions, 
which can be commonly expressed by $\Delta V$,  
play crucial roles in the physical properties of liquids. 

It should be noted that the effects of the anisotropic part 
becomes more and more important with approaching to the liquid-solid phase 
transition, since the system becomes very sensitive to the strength 
and shape of the attractive potential; 
thus, even weak energetic anisotropy can lead to the dramatic effects. 
The dependence $\Delta V$ on $\Omega$ is not necessarily consistent with 
the symmetry favored by the isotropic part $\bar{V}$. 
This causes the energetic frustration, which plays 
key roles in the phase behavior of a supercooled liquid. 

In this Letter, we propose a simple phenomenological 
Landau-type theory based on the coupling 
between two order parameters: (a) density $\rho$ 
determined by the isotropic part ($\bar{S}$) 
of intermolecular attractive interactions 
and (b) bond order $S$ determined by its anisotropic part ($\Delta V$). 
Here the bond parameter $S$ is defined as 
the local number density of molecules having active bonds ({\it i.e.} 
molecules having a local configulation favored by $\Delta V$): 
$S(\vec{r})=\Sigma_i \delta (\vec{r}-\vec{r}_i)$, 
where $\Sigma_i$ is the sum over a unit volume and $\vec{r}_i$ 
is the position vector of a molecule having active bonds $i$.  
We are here aiming at making the simplest model that gives us a  
physical intuition to the phenomena. 
We demonstrate that the frustration between $\rho$ and $S$ is responsible 
for glassification phenomena and the related unusual phenomena 
such as critical-like large-scale density fluctuations 
and phase separation of a supercooled liquid into two liquid (glassy) 
phases. 
The two order parameters are prerequisite for describing the 
physical properties of liquid near its lower stability limit. 

On the basis of this two-order-parameter description, we here construct 
a phenomenological model, focusing only on large-scale fluctuations and the 
associating slow dynamics. 
We average out all the high wavenumber components of $\rho$ and $S$ by 
coarse-graining the system over the length scale much larger 
than the molecular spacing.   
In the standard theory of liquids \cite{Lubensky}, 
the free energy of liquid is given by 
\begin{eqnarray}
F_{\rho}=\int d\vec{r} [\frac{\tau}{2} \delta \rho(\vec{r})^2 
-\frac{a_3}{3} \delta \rho(\vec{r})^3 
+\frac{a_4}{4} \delta \rho(\vec{r})^4], \nonumber 
\end{eqnarray}
where $\tau=a_2(T-T^\ast)$ ($T^\ast$: the lower stability limit of a liquid) 
and $a_i$ is a positive constant. 
By including the gradient terms and 
the coupling between $\rho$ and $S$ into the above 
standard Hamiltonian of liquid, we obtain 
the following Hamiltonian that we believe is relevant to the description 
of liquid: 
\begin{eqnarray}
\beta H&=&F_{\rho}+\int d\vec{r} [\frac{K_\rho}{2}
(\nabla \delta \rho(\vec{r}))^2 
-c_{1 \rho} \delta \rho(\vec{r}) S(\vec{r}) \nonumber \\
&-&c_{1S} \rho(\vec{r}) \delta S(\vec{r})
- \frac{c_{2 \rho}}{2}  \delta \rho(\vec{r})^2 S(\vec{r})  
- \frac{c_{2S}}{2} \rho(\vec{r}) \delta S(\vec{r})^2  \nonumber \\
&+&\frac{\kappa}{2}  \delta S(\vec{r})^2 
+\frac{K_S}{2} (\nabla \delta S(\vec{r}))^2],  
\label{eq:H}
\end{eqnarray}  
where $\beta=1/k_BT$.  $S=\bar{S}+\delta S$, where $\bar{S}$ 
is the average of $S$ and given by $\bar{S}=S_0 \exp(\beta \Delta V)$.  
We neglect a possible tensorial character of $S$ and treat $S$ 
as a scalar. The appearance of $S$ and $\rho$ in the coupling terms 
instead of $\delta S$ and $\delta \rho$ 
is to take into account the fact that that 
they are strongly dependent upon temperature, 
which directly affect the couplings.  
In Eq. (\ref{eq:H}),   
for $c_i>0$ bond formation increases 
density and helps the density ordering, which leads to an increase in 
the ordering temperature, while 
for $c_i<0$ it decreases both of density and the ordering temperature.  
We call the former case the ``positive'' coupling between the order 
parameters and the latter case the ``negative'' coupling. 
As explained later, we believe that this ``negative'' coupling 
between the order parameters is the physical origin 
of glassification phenomena.  

The dynamics of $\delta \rho$ and $\delta S$ is, then, described by  
\begin{eqnarray}
\frac{\partial \delta \rho(\vec{r,t})}{\partial t}
&=&L_\rho \nabla^2 \frac{\delta (\beta H)}
{\delta (\delta \rho(\vec{r,t}))}+ 
\zeta_\rho(\vec{r,t}), \label{eq:kinetic1} \\ 
\frac{\partial \delta S(\vec{r},t)}{\partial t}
&=&-L_S \frac{\delta (\beta H)}
{\delta (\delta S(\vec{r},t))}+ 
\zeta_S(\vec{r},t), \label{eq:kinetic2} 
\end{eqnarray}
where $\zeta_\rho$ and $\zeta_S$ are 
the usual Gaussian noise terms 
and $L_\rho$ and $L_S$ are the bare kinetic coefficients. 
The diffusive equation \cite{Kirkpatrick} likely 
describes 
the dynamics of $\delta \rho$ for a system under the strong damping 
due to large viscosity.  
We propose that Eqs. (\ref{eq:H})-(\ref{eq:kinetic2}) are the 
fundamental equations universally describing slow dynamics of 
supercooled liquids. 

First we consider the density fluctuations in liquids. 
Since the average values of the order parameters are shifted 
by their couplings, we here redefine 
the fluctuations of density and bond order 
parameters around their thermal equilibrium values 
as $\delta \rho^\ast=\delta \rho-<\delta \rho>$ and 
$\delta S^\ast=\delta S-<\delta S>$, where 
\begin{eqnarray}
<\delta \rho>=\frac{c_{1\rho}\bar{S}(\kappa-c_{2S}\bar{\rho})
+c_{1S}(c_{1\rho}+c_{1S})}{(\tau-c_{2\rho}\bar{S})(\kappa-c_{2S}\bar{\rho})
-(c_{1\rho}+c_{1S})^2}, \nonumber \\
<\delta S>=\frac{c_{1S}\bar{\rho}(\tau-c_{2\rho}\bar{S})
+c_{1\rho}(c_{1\rho}+c_{1S})}{(\kappa-c_{2S}\bar{\rho})
(\tau-c_{2\rho}\bar{S})-(c_{1\rho}+c_{1S})^2}. \nonumber 
\end{eqnarray}
Then, the structure factor at small $q$ can be obtained as:
\begin{eqnarray}
&F&(q)=<|\delta \rho^\ast_q|^2> \nonumber \\
&=&\frac{\kappa-c_{2S}\bar{\rho}+K_Sq^2}
{(\tau-c_{2\rho}\bar{S}+K_\rho q^2)(\kappa-c_{2S}\bar{\rho}+K_Sq^2)-
(c_{1\rho}+c_{1S})^2} \nonumber \\
&\cong& \frac{\xi^2}{K_\rho (1+q^2 \xi^2)},  
\label{Fq} 
\end{eqnarray}
where
\begin{equation}
\xi^2 = K_\rho /a_2[T-T^{\ast\ast}(\bar{S})]. \label{xi}
\end{equation}
Here $T^{\ast\ast}(\bar{S}) =T^\ast + (c_{2\rho}/a_2) \bar{S}$. 
In the derivation of the final relation of Eq. (\ref{Fq}), 
we use the weak coupling approximation to neglect the terms 
proportional to $(c_{1\rho}+c_{1S})^2$ in the denominator. 
Equations (\ref{Fq}) and (\ref{xi}) tells us that 
the coupling between two order parameters leads to 
large-scale density fluctuations near the temperature 
$T^{\ast\ast}(\bar{S})$.  
The smallness of the parameter $a_3(\bar{S})$ leads to 
the second-order-like transition 
as in isotropic-nematic transition of 
liquid crystals \cite{deGennesLQ}. 
It is natural to expect that $a_3$ is a decreasing function of $\bar{S}$ 
for the case of $c_i<0$ 
simply because active bonds prevent the formation of a high density phase.  
The decay rate of fluctuations can also be obtained straightforwardly from 
Eq. (\ref{eq:kinetic1}) as 
\begin{equation}
\Gamma_q=\frac{L_\rho q^2}{F(q)}=
\frac{L_\rho K_\rho}{\xi^2}
(1+q^2 \xi^2)q^2. \label{Gamma}
\end{equation} 

Our model indicates that the avoidance of crystallization 
(positional ordering) at $T^\ast$ 
[$T^{\ast\ast}(\bar{S})<T^\ast=T^\ast(0)$] 
and the resulting stabilization 
of a supercooled state of 
liquid are due to the frustration coming from anisotropic 
interactions having non-crystallographic symmetry; and, thus, 
a ``negative'' coupling between $\rho$ and $S$ is a 
necessary condition for good glass formers. 
According to our picture, any liquid that has a ``positive'' coupling 
between $\rho$ and $S$ does crystallize without glassification, 
since crystallization is even induced by bond ordering. 
Thus, we conclude that the way (sign) of coupling between $\rho$ and $S$ 
directly determines  whether molecules just crystallize 
without supercooling or can easily form a glass without 
crystallization. 

Next we consider phase separation below $T^{\ast\ast}(\bar{S})$. 
Because of the coupling between $\rho$ and $S$, 
phase separation can proceed on the $(\rho,S)$ plane in the unstable 
region even for one-component liquid.  
Differently from the phase separation in two component systems, 
the quenching path is automatically determined by 
the shape of the Hamiltonian and the temperature dependence of 
$\bar{\rho}$ and $\bar{S}$. By changing pressure, however, 
we can, of course, change the path. 
For our Hamiltonian, the system is more stable 
along the $S$ axis than along the 
$\rho$ axis. Thus, the phase separation 
takes place rather along the $\rho$ axis initially and then 
the bond order starts to deviate largely from $\bar{S}$. 
Eventually, the system phase separates into two 
phases given by the condition of thermal equilibrium 
$\delta H_{eq}/\delta (\delta \rho)=
\delta H_{eq}/\delta (\delta S)=0$ ($H_{eq}$: 
the equilibrium part of $H$ without the gradient terms). 
Figure \ref{fig:PD} schematically shows how phase separation of 
a one-component liquid can proceed in the $(\rho,S)$ plane. 
The location of the hidden crystal phase is also drawn 
in the figure. 
This unusual phase separation of a single-component liquid 
becomes possible only for the ``negative coupling'' case, since 
the crystallization is avoided even below $T^\ast$. Further, 
the condition that the Vogel-Fulcher temperature $T_0$ is located below 
$T^{\ast\ast}(\bar{S})$ is a prerequisite to this phenomenon.  

Here we discuss the relevance of our simple theory to 
the numerical and experimental findings of some unusual features of liquids. 
First we discuss the anomaly of supercooled water. 
Recently, Stanley {\it et al.} \cite{StanleyS} have proposed 
that there exists a ``second'' critical point (below which 
two liquid phases, a high density liquid and a low-density 
liquid, coexist) for liquid water, on the basis of computer 
simulations using the reliable intermolecular potential. 
Water molecules are known to form hydrogen-bonding network. 
They also proposed a theory that includes the effects of 
hydrogen bonds to the van der Waals equation \cite{StanleyT}. 
In our view, water is one of the most typical examples that have 
the ``negative'' coupling between $\rho$ and $S$. 
This coupling is likely responsible for (i) the large-scale 
critical-like fluctuations in supercooled water and (ii) 
phase separation of water into two water phase having different 
values of $\rho$ and $S$ below $T^{\ast\ast}(\bar{S})$ 
(see Fig. \ref{fig:PD}). 
This is quite consistent with the unusual behavior 
in supercooled water observed in simulations \cite{StanleyS}. 
Without specific assumptions, our model naturally leads to the 
conclusion that density fluctuations have a phase difference of 
180 $^\circ$ to bond order fluctuations: a high density region 
has less bond order, while a low density region has more bond order. 
In water, however, the primary ordering process 
responsible for crystallization is not the density ordering, 
but the bond ordering, in contrast to the cases of 
usual liquids. This makes liquid water quite different from 
other usual molecular liquids. The unique features of water 
associated with the bond ordering nature of crystallization 
will be discussed elsewhere \cite{HTunp}. 

Next we mention the striking experimental evidence 
of phase separation of a supercooled liquid into two liquid phases, 
which was reported by Aasland and McMillan \cite{Aasland}: 
They observed directly the coexistence of two glassy liquids with 
the same composition but different density in a supercooled state 
of Al$_2$O$_3$-Y$_2$O$_3$. 
The similar phenomena have also been reported by many researchers 
(see the references in \cite{Angell,Aasland}). 
These phenomena can also be naturally 
explained by our model, similarly to the above case of water 
(see also Fig. \ref{fig:PD}). 

Further, it should also be noted that there is a number of 
experimental evidence of large-scale fluctuations near 
a glass-transition temperature in several good glass formers, 
which is known as ``Fischer cluster'' \cite{Debye,Fischer2}. 
According to the standard understanding of liquids, there should 
not be large-scale density fluctuations in a single-component 
liquid that are beyond the value determined by the isothermal 
compressibility $K_T$. 
In contrast to this common sense,  
however, Debye reported that long-range density fluctuations 
with a correlation length $\xi_{cluster}$ 
of 200 nm exist in a glassy polymer \cite{Debye}. 
Recently Fischer and his coworkers \cite{Fischer2} 
have found experimentally  
that the large-scale density fluctuations exist near 
the glass-transition temperature $T_g$ commonly 
in several fragile glasses, such as orthoterphenyl, 
poly(methyl methacrylate), and polysiloxane. 
The results strongly suggest that large-scale density fluctuations 
universally exist in various fragile glasses, 
which cannot be understood in the 
framework of the conventional theory of liquids \cite{Fischer2}. 
The structure factor $F(q)$ can be described by the Ornstein-Zernike-type 
one with a correlation length of $\xi_{cluster}$, 
namely, $F(q) \sim 1/[1+(q \xi_{cluster})^2]$ \cite{Fischer2} .  
On the other hand, the decay rate of the fluctuation, $\Gamma_q$, is 
found to depend on the wavenumber $q$ as $q^2$, which is 
suggestive of simple diffusion \cite{Fischer2}. 
This unusual phenomenon observed in supercooled liquids 
can also be naturally explained by our model:  
The Ornstein-Zernike shape of the observed structure factor $F(q)$, 
the growth of the correlation length $\xi_{cluster}$, and 
the diffusive character of the fluctuating slow mode are all 
very much consistent with our predictions   
[see Eqs. (\ref{Fq}), (\ref{xi}), and (\ref{Gamma})]. 

Our model further predicts weaker critical effects 
for a stronger glass former because 
the distance between $T_g$ and  $T^{\ast\ast}(\bar{S})$ 
increases with an increase in the strong nature 
of glass ($\bar{S}$) and, thus, a strong glass former likely 
becomes glassy before approaching to the critical point 
$T^{\ast\ast}(\bar{S})$. 
This is consistent with the fact that the large-scale density 
fluctuations have so far been observed only in rather fragile glasses.

In conclusion, we propose a simple universal picture of 
liquids near the liquid-solid phase transition. 
We demonstrate that the energetic frustration 
between the isotropic and anisotropic parts of 
intermolecular attractive interactions 
(the ``negative'' coupling between the density 
and bond order parameter) 
is a prerequisite to the physical description of supercooled liquids. 
Our simple Ginzburg-Landau-type theory based on the two-order-parameter 
description provides us with intuitive physical explanations 
for the unusual phenomena of supercooled liquids: (i) large-scale density 
fluctuations in supercooled liquids (``Fischer cluster'') \cite{Fischer2}, and 
(ii) phase-separation phenomena of a one-component supercooled liquid into 
two liquid (glassy) phases \cite{Aasland,StanleyS}. 

The author is grateful to S.F. Edwards, E.M. Telentjev, C.-Y.D. Lu, 
and D.V. Grinev for valuable discussions and comments. 
He also thanks to E.W. Fischer, S. Dietrich, and R. Evans for 
fruitful discussions. 
This work is partly supported by a Grant-in-Aid 
from the Ministry of Education, Science, Culture, and Sports, Japan.

\vspace{-0.5cm}
\begin{minipage}{8cm}
\begin{figure}

\begin{center}
\psfig{figure=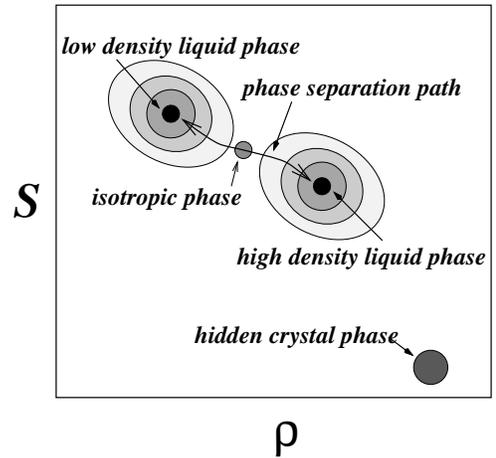,width=6cm,angle=-90}
\end{center}
\caption{Schematic figure of the phase behavior of a 
supercooled liquid on the $(\rho,S)$ plane below 
a spinodal line.  A homogeneous one-component liquid 
can demix into two phases having different $\rho$ and $S$. } 
\label{fig:PD}
\end{figure}
\end{minipage}
\makebox[2cm]{}
\end{multicols}
\end{document}